\documentclass[useAMS,usenatbib]{mn2e}
\usepackage{graphicx}
\title[On rapid migration and accretion within disks around supermassive black holes]{On rapid migration and accretion within disks around supermassive black holes}

\author[B. McKernan, K.E.S. Ford, W.Lyra, H.B. Perets, L.M.Winter \& T.Yaqoob]{B. McKernan$^{1,2,3}$\thanks{E-mail:bmckernan at amnh.org (BMcK)}, K.E.S. Ford$^{1,2,3}$, W. Lyra $^{2}$, H.B.Perets$^{4}$, L.M.Winter$^{5,7}$ \& T. Yaqoob$^{6}$ \\
$^{1}$Department of Science, Borough of Manhattan Community College, City University of New York, New York, NY 10007\\
$^{2}$Department of Astrophysics, American Museum of Natural History, New York, NY 10024\\
$^{3}$Graduate Center, City University of New York, 365 5th Avenue, New York, NY 10016\\
$^{4}$Harvard-Smithsonian Center for Astrophysics, 60 Garden St., Cambridge, 
MA02138\\
$^{5}$Center for Astrophysics \& Space Astronomy, University of Colorado, Boulder, CO 80303\\
$^{6}$Department of Physics \& Astronomy, Johns Hopkins University, Baltimore, MD 21218\\
$^{7}$Hubble Fellow\\
}
\begin{document}

\date{Accepted. Received; in original form}

\pagerange{\pageref{firstpage}--\pageref{lastpage}} \pubyear{2008}

\maketitle

\label{firstpage}

\begin{abstract}
Galactic nuclei should contain a cluster of stars and compact objects in the 
vicinity of the central supermassive black hole due to stellar evolution, 
minor mergers and gravitational dynamical friction. By analogy with 
protoplanetary migration, nuclear cluster objects (NCOs) can migrate in the 
accretion disks that power active galactic nuclei by exchanging angular 
momentum with disk gas. Here we show that an 
\emph{individual} NCO undergoing runaway outward migration comparable to Type 
III protoplanetary migration can generate an accretion rate corresponding to 
Seyfert AGN or quasar luminosities. Multiple migrating NCOs in an AGN disk can 
dominate traditional viscous disk accretion and at large 
disk radii, ensemble NCO migration and accretion could provide sufficient 
heating to prevent the 
gravitational instability from consuming disk gas in star formation. The 
magnitude and energy of the X-ray soft excess observed at $\sim 0.1-1$keV in 
Seyfert AGN could be explained by a small population of $\sim 10^{2}-10^{3}$ 
accreting stellar mass black holes or a few ULXs. NCO migration and accretion 
in AGN disks are therefore extremely important mechanisms to add to realistic 
models of AGN disks. 

\end{abstract}

\begin{keywords}
galaxies: active --
galaxies: individual -- galaxies: LINERs -- techniques: spectroscopic
           -- X-rays:  line -- emission: accretion -- disks :galaxies
\end{keywords}

\section{Introduction}
\label{sec:intro}
Gigantic gas disks feeding supermassive black holes are prone to the 
gravitational instability beyond $\sim 10^{3}$ Schwarzschild radii 
\citep[e.g.][]{b46,b44}. In order to maintain large-scale gas disks 
implied by huge accretion rates for AGN lifetimes, additional heating of the 
outer disk is required \citep{b22}, but a 
universal heating mechanism has yet to be found \citep[e.g.][]{b23}. The 
converse of this problem is that in low luminosity AGN, the implied low 
accretion rates \citep{b4} should make it impossible to maintain 
an inflated, obscuring torus in these objects \citep{b3} via 
radiation pressure from the central engine alone \citep{b15}. 

Here we discuss the interaction of the cluster of stars and compact 
objects in galactic nuclei with AGN accretion disks. Nuclear cluster 
objects (NCOs) with orbits coincident with the disk can accrete strongly and 
exchange angular momentum with disk gas and each other. By migrating and 
accreting within the disk, NCOs can play a role in the problem of disk 
heating and the inferred AGN accretion rate. The interaction of NCOs 
with AGN disks has long been discussed \citep[e.g.][]{b57,b58}, while 
NCO migration within AGN disks \citep[e.g.][]{b51,b44,b45} and NCO accretion 
\citep[e.g.][]{b35,b98,b99} must play a role in models of the AGN central 
engine. In this Letter, we emphasize the importance of runaway NCO migration 
\emph{within} AGN disks due to the torque produced by gaseous fluid elements 
in the disk as they perform horseshoe U-turns near the NCO. We point out that 
an \emph{individual} NCO migrating rapidly due to co-orbital corotation torque 
\citep{b28} can drive accretion rates that \emph{dominate} the viscous disk 
accretion rates believed to power AGN. We also point out that NCO accretion 
can account for the soft X-ray excess observed in Seyfert AGN and the 
inflation of the accretion flow at large radii even when radiation pressure 
from the central engine is very low. We conclude that realistic models of AGN 
central engines must include a population of migrating and accreting NCOs.

\section{NCOs and the analogy with protoplanetary disks}
\label{sec:migration}
The largest, supermassive, black holes in the Universe 
($M_{\rm BH}\sim 10^{6}-10^{9}M_{\odot}$) live in galactic centers 
\citep[e.g.][]{b5}. These behemoths are expected to be closely surrounded by a
 dense nuclear cluster of objects as a result of stellar evolution, dynamical 
friction, secular evolution and minor mergers \citep[e.g.][]{b12,b11,b8}. In 
our own Galactic nucleus, the implied distributed mass 
within $\sim$1pc is $\sim 10\%-30\%$ of the mass of SgrA* itself 
\citep{b17}. Mass segregation may play a role in determining the NCO 
population \citep[e.g.][]{b1,b8}, but we should expect $\sim 10^{6}$
 NCOs of average mass $\sim 1M_{\odot}$ within 1pc of a SgrA* black hole.
In a given nucleus, the exact proportion of stellar and compact NCOs will 
depend on the history of the nucleus host galaxy \citep[e.g.][]{b98}. If a 
large quantity of gas arrives in the innermost pc of a galactic nucleus, 
it will likely lose angular momentum and accrete onto the central supermassive
 black hole, but in doing so must also interact with the NCO population. 
Depending on the aspect ratio of the disk that forms, a few percent 
of NCOs orbits are likely to coincide with the accretion flow. A small 
percentage of NCOs coincident with the disk could lead to few$\times 10^{4}$ 
NCOs in the accretion flow around a SgrA* black hole. While multi-object 
migration in disks is an open problem beyond the scope of this Letter, the 
existence of \emph{one} migrator can have important consequences for the 
observational appearance of galactic nuclei. Given the number of NCOs 
available, it is probable that the effects described below are important in 
many AGN disks.

By analogy 
with protoplanetary disk theory, and following e.g. \citet{b44,b45}, we
 assume that NCOs in thin AGN disks exchange angular momentum with disk gas 
and migrate. This assumption is not unreasonable since the mass ratios 
involved, disk densities and related physics (gravitational 
instability and magneto-rotational instability (MRI)) are similar for AGN and 
protoplanetary disks at large radii, although temperatures differ 
substantially at small radii. AGN disks at small radii are relativistic (both 
Special and General). However, 
fully general relativistic global MHD disks appear globally similar to their 
Newtonian counterparts until close to the radius of last stable orbit 
($R_{isco}$) \citep[e.g.][]{b77}. Likewise, self-gravity in AGN 
accretion disks will be important \citep[e.g.][]{b46,b51}. However, disk 
self-gravity does not appear to have a major effect on the
 torques involved in satellite migration, except possibly accelerating 
migration outwards \citep{b27}. Obviously, simulations are required to 
understand the parallel with AGN disks in detail and we will return to this in 
future work. 

The aspect of protoplanetary disk theory that most concerns us here is that 
dense satellites perturb disk symmetry, suffer a net torque and exchange angular 
momentum with disk gas \citep[see e.g.][and references 
therein]{b20,b19,b24}. Even where the disk is self-gravitating 
\citep[e.g.][]{b27}, or ionized enough to maintain MRI \citep[e.g.][]{b13}, the 
torque can operate very effectively in exchanging angular momentum between the 
satellite and the gas. As a result, the satellite can migrate within the gas 
disk, sometimes very rapidly indeed. In classical protoplanetary disk theory, 
Type I migration (small migrator, $q \leq 10^{-5}$) occurs on a timescale of 
\citep{b24}
\begin{equation}
\tau_{\rm I}=\frac{1}{N}\frac{M}{q\Sigma r^{2}} \left( \frac{h}
{r}\right)^{2}\frac{1}{\omega}
\label{eq:type1}
\end{equation}
where $M$ is the central mass, $q$ is the ratio of the satellite mass to the 
central mass, $\Sigma$ is the disk surface density, $h/r$ is the disk aspect 
ratio and $\omega$ is the satellite angular frequency. The numerical factor 
$N$ depends on the ratio of radiative to dynamical timescales and is a 
function of the power-law indices of $\Sigma, T$ and entropy \citep{b19,b24}.
 Larger (Type II) migrators ($q > h^{3}$) can open gaps in the disk and 
migrate inwards on the viscous accretion timescale of the disk, 
$\tau_{\alpha}$, (i.e. $\tau_{\rm II}=\tau_{\alpha}$). An intermediate form 
of migration (Type III) can occur where the migrator ($q \sim 10^{-5}-10^{-4}$) 
perturbs the disk, but not enough to open a gap \citep[e.g.][]{b28,b25}. This 
form of migration can occur very quickly in either direction and may in fact 
runaway outwards depending on the flow of gas around the migrator \citep{b26}.
Type III migration occurs on approximately the dynamical (orbital) timescale 
of the disk ($\tau_{\rm III} \sim \tau_{\omega}$).

\section{AGN accretion and NCOs}
\label{sec:agns}
Accretion onto the central black hole in AGN occurs via the transfer of 
angular momentum between parcels of gas in a disk. A mass of gas moving 
inwards ($m_{\rm in}$) in the disk will drive a mass of gas outwards 
($m_{\rm out}$), conserving angular momentum so that $m_{\rm in}/m_{\rm out} 
\sim \sqrt{R_{\rm out}/R_{\rm in}}$ for a Keplerian disk. In the case of the 
classic stationary thin disk, the mass flux through the disk around a 
Schwarzschild black hole (and inferred accretion rate) is given by \citep{b66}
 \begin{equation}
\dot{M}=\frac{2\pi \alpha \Sigma c_{s}^{2}}{\omega \left[ 
1-\left(\frac{R_{isco}}{R}\right)^{1/2}\right]}=\frac{\Delta M}{\Delta t}
\label{eq:shakura}
\end{equation}
where $\alpha$ is the viscosity parameter, $c_{\rm s}$ is the sound speed and 
$\Delta t$ is the viscous timescale 
of the disk, given by $\tau_{\alpha} \sim 
\alpha^{-1}(h/r)^{-2} \tau_{\omega}$. Friction in the viscous 
disk generates luminosity which is parameterized as $\eta \dot{M} c^{2}$ where
 $\eta$ is the efficiency of gravitational energy 
release ($\eta \sim 0.06$ for a Schwarzschild metric \citep{b66} up to 
$\sim 0.42$ for a maximally spinning Kerr black hole) for a 
standard thin disk. Heat release per unit disk area for a thin disk in a 
Keplerian potential is $\propto \dot{M}/r^{3}$ \citep[e.g.][]{b39}.

Now, if we substitute the outflowing gas parcel ($m_{\rm out}$) above with an 
out-migrating NCO of identical mass, $m_{\rm in}$ will be identical, but the 
inflow timescale will now be the NCO migration timescale ($\tau_{\rm m}$) 
rather than the viscous disk timescale ($\tau_{\alpha}$). The disk luminosity 
and heating generated by the inflow is still produced by the same 
underlying microphysics (i.e. gas friction in the disk). The only difference 
is that $\Delta t$ in equation~\ref{eq:shakura} is now the migration 
timescale. If $\tau_{\rm m} \ll \tau_{\alpha}$ then $\dot{M}$ due to 
migration can dominate viscous accretion.

Figure~\ref{fig:dt} shows the relevant timescales as a function of 
radius, using a thin accretion disk surrounding a 
$10^{8}M_{\odot}$ supermassive black hole with $\dot{M}$ constant (at 0.5 
Eddington) as a function of radius \citep{b22}. $\tau_{\rm dyn}$ is the 
dynamical (orbital) timescale and the relevant timescale for Type III 
migration and $\tau_{Q} \sim \alpha^{-1} \tau_{\rm dyn}$ is the local disk 
heating timescale. $\tau_{\rm I}$ has been calculated from eqn.~\ref{eq:type1}
 for a $q=10^{-6}$ ($10^{2}M_{\odot}$) NCO migrator in the adiabatic limit 
using $\Sigma^{-2}$, $T \propto r^{-3/4}$ from \citet{b22}. $\tau_{\rm I}$ 
truncates at $\sim 500r_{g}$ from eqn.~\ref{eq:radius} below and the kink in 
$\tau_{\rm I}$ comes from model parameters of the disk in \citet{b22}.

\subsection{NCO migration in AGN disks}
\label{sec:ncos}
For an NCO to actually migrate within the gas disk, the gas in a ring 
co-rotating with the NCO should have an angular momentum comparable to that of 
the NCO. This condition corresponds to
\begin{equation}
2\pi\Sigma r \Delta r \sim m
\end{equation}
where $\Sigma$ is the disk surface density, $r$ is the distance of the NCO from 
the supermassive black hole, $\Delta r$ is the width of the co-rotating gas 
region (roughly the Hill sphere radius) and $m$ is the mass of the NCO. 
The radius of the Hill sphere for a 
compact object on a circularized orbit ($e=0$) is $r_{\rm H}\approx rq^{1/3}$ 
where $r$ is the orbital radius and q is the mass ratio of the compact object to
 the central black hole. 
Assuming $\Delta r \sim r q^{1/3}$, the innermost disk radius (in units of 
gravitational radii) at which migration can occur is
\begin{equation}
R^{in}(r_{g}) \sim \frac{10^{3}m_{\odot}^{1/2}}
{q^{1/6}M_{8}\Sigma^{1/2}}r_{g}
\label{eq:radius}
\end{equation}
where $m_{\odot}$ is the mass of the migrator in units of $M_{\odot}$, 
$M_{8}$ is the mass of the supermassive black hole in units of 
$10^{8}M_{\odot}$, $q=m_{\odot}/M_{8}$ and $\Sigma$ is the surface 
density of the AGN disk in units of g $\rm{cm}^{-2}$. 

\begin{figure}
\includegraphics[width=3.35in,height=3.35in,angle=-90]{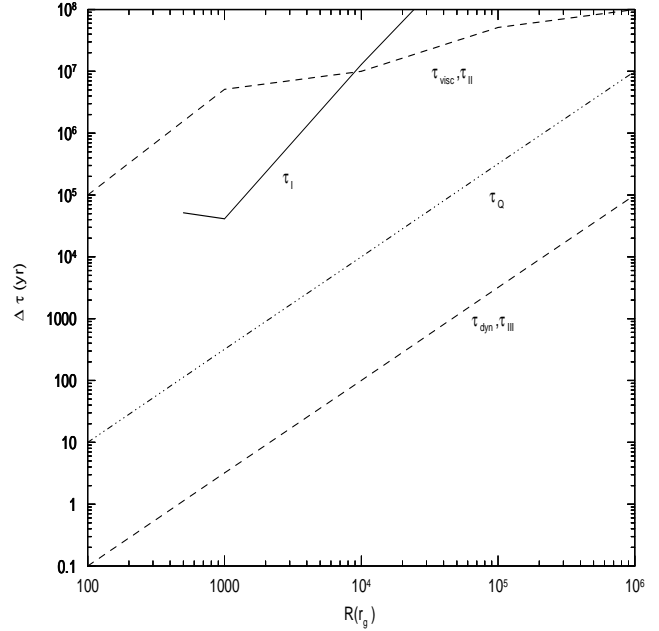}
\caption{Timescales as a function of radius for a standard thin accretion 
disk ($\alpha \sim 0.01, (h/r) \sim 10^{-2}$) around a 
$10^{8}M_{\odot}$ supermassive black hole. $\tau_{\rm dyn}=\tau_{\omega}$ is 
the dynamical/orbital timescale and the timescale on which Type III migration 
occurs, $\tau_{Q}$ is the typical heating timescale, $\tau_{I}$ is the 
timescale on which Type I migration can occur in the disk for a 
$10^{2}M_{\odot}$ NCO and $\tau_{\rm visc}=\tau_{\alpha}$ is the viscous 
(accretion) timescale as calculated from the disk model of \citet{b22}.
\label{fig:dt}}
\end{figure}

Most NCOs will undergo Type I migration ($q <10^{-5}$) in thin AGN disks. 
Exceptions include OB stars and stellar mass black holes around 
$10^{6}M_{\odot}$ black holes, which at $q \sim 10^{-5}-10^{-4}$ could 
undergo rapid Type III migration in the disk. Around 
$10^{8}M_{\odot}$ black holes, only intermediate mass black holes 
$>10^{3}M_{\odot}$ could undergo rapid Type III migration. Type II migration 
by IMBHs or supermassive stars with $q>10^{-4}$ could open gaps in thin AGN 
disks \citep[e.g.][]{b44}. Gaps in an accretion disk would truncate NCO 
migration, possibly leading to three-body interactions with IMBHs. Multiple 
IMBHs in an AGN disk would show up as deficits in broad spectral lines 
that vary collectively on a timescale of $\tau_{\alpha}$. 

\begin{figure}
\includegraphics[width=3.35in,height=3.35in,angle=-90]{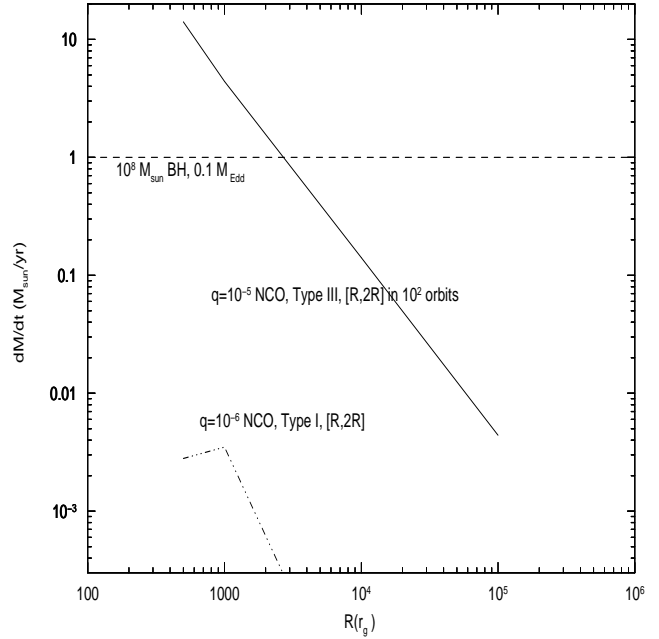}
\caption{dM/dt as a function of radius for the accretion disk around a 
$10^{8}M_{\odot}$ supermassive black hole in 
Fig.~\ref{fig:dt} above. Following \citet{b28}, we plot the corresponding 
$\dot{M}$ due to the Type III migration of \emph{one} $q=10^{-5}$ 
($10^{3}M_{\odot}$) 
NCO whose radius doubles over $\sim 10^{2}$ orbits. For Type I migration, we 
assume \emph{one} $q=10^{-6}$ ($10^{2}M_{\odot}$) NCO, whose radius doubles over 
the migration timescale.
\label{fig:dMdt}}
\end{figure}

Figure~\ref{fig:dMdt} shows $\dot{M}$ as a function of radius for the 
accretion disk from Fig.~\ref{fig:dt}. The model disk of \citet{b22} assumes 
constant $\dot{M}$ with radius, so we show the $10\%$ Eddington mass accretion 
rate for a $10^{8}M_{\odot}$ as a dashed line. The solid line represents a 
$q=10^{-5}$ NCO undergoing rapid Type III migration and doubling its radius in
 $\sim 10^{2}$ orbits, following the simulated $\sim 50\%$ increase in radius 
in a few $10$'s of orbits for protoplanetary satellites \citep[e.g.][]{b28}. 
The dash-dot line 
represents a $q=10^{-6}$ NCO undergoing Type I migration and doubling its 
radius on the Type I migration timescale. We note several important points 
from Fig.~\ref{fig:dMdt}. First, the runaway migration of a \emph{single} 
$q=10^{-5}$ NCO near $\sim 10^{3}r_{g}$ can generate the Eddington mass 
accretion rate in an AGN, corresponding to quasar-like luminosity. Based on 
the environment of SgrA*, up to a few$\times 10^{7}$ NCOs could lie 
within 1pc of the $10^{8}M_{\odot}$ in Fig.~\ref{fig:dMdt}, leading to a 
large NCO population in the accretion disk. Second, since runaway migration is 
greatest for large disk surface density \citep{b28}, runaway NCO migration is 
most likely to occur deeper in the AGN potential where it can generate very 
high $\dot{M}$. Third, if the speed or magnitude of runaway migration is 
greater for NCOs in AGN disks than in \citep{b28}, $\dot{M}$ in 
Fig.~\ref{fig:dMdt} could be substantially higher. The local disk heating 
($\propto \dot{M}/r^{3}$) associated with runaway migration within a few
 thousand $r_{g}$ exceeds local viscous heating for Seyfert AGN 
($\sim 0.1 L_{\rm Edd}$) by a factor of a few. If $\dot{M}$ due to 
viscous accretion drops with radius, local 
heating due to migration at large radii ($\propto \dot{M}/r^{3}$) could 
dominate viscous heating, particularly since the NCO population should 
increase as the radius squared. Note that, if the NCO migrates 
\emph{inwards} rather than outwards, the Type III and Type I migration curves 
in Fig.~\ref{fig:dMdt} will be lower by a factor of $\sqrt 2$ since 
$m_{\rm in}/m_{\rm out} = \sqrt{R_{\rm out}/R_{\rm in}}$ for a Keplerian disk.
 Interestingly, an NCO that is not tidally disrupted while accreted onto the 
black hole should correspond to a low luminosity ($\eta \dot{M} c^{2}$) mode 
of accretion.

\subsection{Clearing out \& tidal disruption}
\label{sec:clear}
The very large initial number of NCOs expected in the AGN disk should lead to 
a rapid clearing out of NCOs due to three-body interactions. Naively we might 
expect roughly half the cleared out NCOs to be ejected from the disk and half 
to be accreted onto the supermassive black hole. The clearout duration 
($\tau_{\rm clear}$) could constitute a very high accretion rate ($\sim 0.5 
N_{\rm NCO}/\tau_{\rm clear}$) onto the supermassive 
black hole. If the clearout occurs e.g. on the average dynamical timescale at 
$10^{6}r_{g}$ ($\sim 0.1$Myr) in Fig.~\ref{fig:dt}, for $\sim 10^{6}$ NCOs in 
the disk surrounding a $10^{8}M_{\odot}$ black hole, the NCO accretion rate is 
$\dot{M}_{\rm Edd}$ for $\sim 0.1$Myr. Compact NCOs are less effected by tidal
 shearing and so may represent the most radiatively inefficient mode of 
accretion while tidally disrupted stellar NCOs will contribute to the AGN 
luminosity ($\eta \dot{M} c^{2}$) during this phase. After clearing out, a 
small population of migrators should be left in the thin AGN disk. However, 
gradual orbital decay of NCOs intersecting the disk must occur over time
\citep[e.g.][]{b57,b51,b70}. Phenomena such as resonant relaxation 
\citep{b88} and the Kozai mechanism \citep[e.g.][]{b69,b59} will contribute to 
NCO capture by the disk, although on timescales longer than the AGN 
disk lifetime ($\sim 10$Myrs) in most cases. The capture rate for NCOs on 
small radius 
orbits is the dominant effect over the AGN lifetime \citep[e.g.][]{b51}, so 
$\sim 10^{3-4}$ NCOs could be added to the disk in Fig.~\ref{fig:dMdt} within 
$\sim 10^{5}r_{g}$ over a $10$Myr disk lifetime. 

\section{Accretion onto NCOs}
\label{sec:observations}
NCOs can have very different appearances depending on their accretion rates 
from the disk. The most massive stellar NCOs, accreting rapidly from the disk,
 can become luminous blue, Wolf-Rayet stars \citep[e.g.][]{b51}. Minor black 
holes accreting rapidly can look like high mass X-ray binaries or ULXs 
depending on $M_{BH}$ \citep[see e.g.][]{b98,b99}. Neutron stars
 accreting at a high rate can spin up rapidly becoming pulsars, contributing to
(quasi) periodic signals in the radio and X-ray bands. Accreting white dwarfs 
will heat up and contribute to the overall IR flux and generate Type Ia 
supernovae. If the mass of the accreting NCO approaches $q \sim 10^{-5}$, the 
NCO may begin to migrate very rapidly in the disk (see above). For 
NCOs not in the disk, sporadic accretion will occur onto compact NCOs punching
 through the disk on orbits on small fractions of their orbital timescales, 
contributing to observed volume filling broad-line and narrow-line 
emission \citep[e.g.][]{b56}. 

\citet{b35} point out that a stellar mass black hole accreting from a dense 
gas disk can accrete at near Eddington rates($\dot{M}_{\rm Edd}$). By 
contrast, strong magnetic 
fields may inhibit accretion onto neutron stars and white dwarfs 
\citep{b36}. For a population of $\sim 10^{4}$ NCOs in the disk 
around a SgrA* black hole, with fiducial compact NCO population of \citep{b1}: $1\%$(black holes, accreting at $\dot{M}_{\rm Edd}$), and 
$3\%$(neutron stars) and $20\% $(white dwarfs) accreting at 
$0.1\dot{M}_{\rm Edd}$, the total compact NCO luminosity contribution 
corresponds to $\sim 10^{-3} \dot{M}_{\rm Edd}$ of the central supermassive 
black hole. Therefore, accretion onto compact NCOs could dominate viscous 
disk accretion in lower luminosity AGN and LINERs \citep[e.g.][]{b4}. A single 
10$M_{\odot}$ black hole accreting at $\sim \dot{M}_{\rm Edd}$ located 
$\sim 1$pc from a $10^{42}$ erg/s AGN ionizing continuum will dominate the 
local ionization field within $\leq 0.05$pc. A population of 
$\sim 10^{2}$ such NCO accretors distributed around $\sim 1$pc would 
dominate low luminosity AGN heating of the outskirts of the disk. 

A soft X-ray bump (the 'soft excess') of $\sim 10^{40-41}$ 
erg $\rm{s}^{-1}$ and uncertain origin is observed in many Seyfert AGN, with constant temperature over four orders of magnitude in black hole mass 
\citep[e.g.][]{b67,b55,b97}. The magnitude and constant temperature of the 
soft excess could be explained either by $\sim 10^{2}-10^{3}$ stellar 
mass black holes accreting like Galactic high-mass X-ray binaries 
\citep[e.g.][]{b68} or by a handful of ULXs \citep{b99}. If there is a 
correlation 
between the soft excess and powerlaw luminosities \citep{b54}, there may be a 
link between NCO accretion rate and accretion onto the supermassive black 
hole. Intrinsic soft excess variability will depend on the 
number of NCO accretors. In the limit of a single highly luminous ULX with 
mass ratio $q$, the soft excess will vary on timescales $q\tau_{pl}$, where 
$\tau_{pl}$ is the timescale of variation of the powerlaw component. In the 
limit of a large number of accreting NCOs, the \emph{magnitude} of the 
intrinsic soft excess luminosity variability will be much smaller than 
powerlaw variability.

The heat liberated by migration ($\propto \dot{M}/r^{3}$) drops quickly with 
increasing radius. Unless the migration is rapid (large $\dot{M}$), heating 
due to NCO accretion should dominate at large radii, since the number of NCOs 
should grow as the square of the radius. A 
large enough number of accreting compact objects (black holes and neutron 
stars) can generate sufficient hard radiation/radiation pressure to inflate 
the accretion flow at large radii, \emph{regardless of the rate of accretion 
onto 
the supermassive black hole}, so we should expect inflated accretion flows in 
even low luminosity galactic nuclei. The Type I migration timescale should 
increase as $(h/r)^{2}$, so migration should become inhibited at large radii, 
unless NCOs accrete enough so that $q>10^{-5}$ and rapid Type III migration 
can occur.

\section{Conclusions and Future Work}
\label{sec:conclusions}
A cluster of stars and compact objects is expected in the central pc of most 
galactic nuclei. The exchange of angular momentum between an accretion disk 
of gas and nuclear cluster objects (NCOs) can lead to NCO migration and NCO 
accretion. Here we emphasize that runaway outward migration of NCOs in AGN 
disks, similar to Type III migration in protoplanetary disks, generates
accretion rates that can dominate traditional viscous disk accretion rates. 
Disk heating due to NCO migration/accretion can inflate the accretion flow 
and counter the gravitational instability at large radii. Inflation of the 
accretion flow due to compact objects in the 
disk may account for obscuring structures in LINERs and low luminosity AGN. 
A small population of accreting stellar mass black holes can account for the 
magnitude and constant temperature of the soft X-ray excess observed in 
Seyfert AGN. 

We also note some important implications for future work: 1) A large 
population of migrating NCOs in a gas disk could help close the final parsec 
gap in stalled supermassive black hole binary mergers \citep[e.g.][]{b7} by 
exchanging angular momentum with the SMBH. 2) NCO clear-out from AGN disks 
may yield a population of high velocity NS and WD in galactic bulges and 
haloes, 
together with high velocity stars \citep[e.g.][]{b31}. High velocity pulsars 
in our own Galaxy have velocities similar to escape speeds from around Sgr A* 
\citep[e.g.][]{b32}. 3) Differences between NCO populations in different 
nuclei could account for observed differences in AGN in spiral and 
elliptical hosts. 4) IMBHs build up mass most 
efficiently as NCOs in an AGN disk and ULX signatures should be observed in 
galactic nuclei \citep{b99}. 5) Radiatively inefficient, thick 
($h/r >0.1$) disks may drive more out-migration than thin-disks. 6) NCO 
migration in a disk threaded with magnetic fields can drive MHD outflows.

We conclude that realistic models of AGN disks should include 
populations of migrating and accreting NCOs.
\section*{Acknowledgements}
We acknowledge useful discussions with Mordecai Mac Low, Sijme-Jan 
Paardekooper and Clement Baruteau on protoplanetary disk theory.

%\bsp

\label{lastpage}

\end{document}